\documentclass[final,twocolumn]{revtex4}

\usepackage{epsfig,graphicx}
\usepackage{amssymb}
\usepackage{amsfonts}
\usepackage{bm}

\def\fun#1#2{\lower3.6pt\vbox{\baselineskip0pt\lineskip.9pt
\ialign{$\mathsurround=0pt#1\hfil ##\hfil$\crcr#2\crcr\sim\crcr}}}

\newcommand{\beq}{\begin{eqnarray}}
\newcommand{\eeq}{\end{eqnarray}}
\newcommand{\be}{\begin{equation}}
\newcommand{\ee}{\end{equation}}

\def\fun#1#2{\lower3.6pt\vbox{\baselineskip0pt\lineskip.9pt
\ialign{$\mathsurround=0pt#1\hfil ##\hfil$\crcr#2\crcr\sim\crcr}}}

\newcommand{{\SD}}{\rm SD}

\newcommand{\vep}{\mbox{\boldmath${\it p}$}}

\newcommand{\ver}{\mbox{\boldmath${\it r}$}}
\newcommand{\ves}{\mbox{\boldmath${\it s}$}}

\begin{document}

\title{The Hyperfine Splittings in Bottomonium and the $B_q (q=n,s,c)$
Mesons}

\author{\firstname{A.M.}~\surname{Badalian}}
\email{badalian@itep.ru}
\affiliation{Institute of Theoretical and Experimental Physics, Moscow, Russia}

\author{\firstname{B.L.G.}~\surname{Bakker}}
\email{blg.bakker@few.vu.nl} \affiliation{Department of Physics
and Astronomy, Vrije Universiteit, Amsterdam, The Netherlands}

\author{\firstname{I.V.}~\surname{Danilkin}}
\email{danilkin@itep.ru} \affiliation{Institute of Theoretical and
Experimental Physics, Moscow, Russia}\affiliation{Gesellschaft fur
Schwerionenforschung (GSI) Planck Str. 1, 64291 Darmstadt,
Germany}


\begin{abstract}
A universal description of the hyperfine splittings (HFS) in
bottomonium and the $B_q\; (q=n,s,c)$ mesons is obtained with a
universal strong coupling constant $\alpha_s(\mu)=0.305(2)$ in a
spin-spin potential. Other characteristics are calculated within
the Field Correlator Method, taking the freezing value of the
strong coupling independent of $n_f$. The HFS $M(B^*)-
M(B)=45.3(3)$~MeV, $M(B_s^*) - M(B_s)=46.5(3)$~MeV are obtained in
full agreement with experiment both for $n_f=3$ and $n_f=4$. In
bottomonium, $M(\Upsilon(9460))- M(\eta_b)=70.0(4)$~MeV for
$n_f=5$ agrees with the BaBar data, while a smaller HFS, equal to
64(1)~MeV, is obtained for $n_f=4$. We predict HFS
$M(\Upsilon(2S))-M(\eta_b(2S))=36(1)$~MeV, $M(\Upsilon(3S))-
M(\eta(3S))=27(1)$~MeV, and $M(B_c^*) - M(B_c)= 57.5(10)$~MeV,
which gives $M(B_c^*)=6334(1)$~MeV, $M(B_c(2\,{}^1S_0))=6865(5)$~MeV, 
and $M(B_c^*(2S\,{}^3S_1))=6901(5)$~MeV.
\end{abstract}

\maketitle

\section{Introduction}

Recently, $\eta_b(1S)$ has been discovered by the BaBar collaboration in
the radiative decays $\Upsilon(3S)\rightarrow \gamma \eta_b(1S)$
\cite{ref.1} and $\Upsilon(2S)\rightarrow \gamma \eta_b(1S)$
\cite{ref.2}, with a mass (averaged over two results)
$M(\eta_b)=9391.1\pm 3.1$~MeV. It gives a rather large hyperfine
splitting (HFS), $\Delta(b\bar b)=
M(\Upsilon(1S))-M(\eta_b(1S))=69.9\pm~3.1$~MeV. Later this mass
was confirmed by the CLEO collaboration also in the radiative
$\Upsilon(3S)\rightarrow \gamma\eta_b(1S)$ decay \cite{ref.3}.
This important new information allows to test again our
understanding of the hyperfine (HF) interaction in QCD.

Although a spin-spin potential between  heavy quarks was used in
numerous studies, the parameters defining this potential significantly
differ in different models. As a result, theoretical predictions for
the mass difference $\Delta(b\bar b)= M(\Upsilon(9460))-M(\eta_b(1S))$
vary in a wide range: $35-90$~MeV \cite{ref.4}--\cite{ref.9} and in most
cases they are smaller than the experimental number. On a fundamental
level the spin-spin potential $V_{ss}({\rm lat})$ has been studied in
detail in quenched QCD (see \cite{ref.10} and references therein). This
lattice potential appears to be compatible with zero at distances $r\geq
0.30$~fm and (for unknown reasons) has negative sign at smaller $r$ (with
a large magnitude); in any case the lattice potential does not contradict
the Fermi-Breit potential with $\delta^3(\ver)$ \cite{ref.11}, although the
behavior of the spin-spin potential at $r\leq 0.3$ fm remains uncertain.

On the other hand, a detailed phenomenological analysis given in
Ref.~\cite{ref.4} has demonstrated the importance of the smearing of
the $\delta^3(\ver)$-function, from which one may expect that for
heavy mesons, containing a $b$-quark, the use of $\delta^3(\ver)$
may be a good approximation. For lighter mesons, like $D$, $D_s$, and
charmonium, a nonperturbative spin-spin potential may be essential,
giving a contribution $\sim 10\%$ \cite{ref.12}. Here we concentrate
on bottomonium and the $B_q~(q=n,s,c)$ mesons, for which nonperturbative
contributions are small, and neglect the smearing effect, in this way
avoiding to introduce several unknown parameters.

Our main goal here is the extraction of the strong coupling
$\alpha_{\rm hf}(\mu)$ from known HFS. In theoretical models two
typical choices of $\alpha_{\rm hf}$ are used:

\begin{enumerate}

\item In the first one, ``a universal" $\alpha_{\rm hf}$ is used.
For example, in Ref.~\cite{ref.5}  $\alpha_{\rm hf}=0.36$ taken,
was obtained from a fit to the mass difference $M(J/\psi)-
M(\eta_c(1S))=117$~MeV, but their HFS, $M(\Upsilon(9460))-
M(\eta_b) = 87$~MeV, is $\sim 25\%$ larger than the experimental
number. In Ref.~\cite{ref.9}, using a smaller $\alpha_{\rm
hf}=0.339$ a good description of the HFS of the $B$ and $B_s$
mesons was obtained. However, a comparison of their and our
results is difficult, because a large string tension,
$\sigma=0.257$~GeV$^2$, was taken in \cite{ref.9}, while here and
in Ref.~\cite{ref.4} the conventional value $\sigma=0.18$~GeV$^2$
is used.

\item
The second choice, with a scale $\mu$ dependent on the quark mass,
is mostly used in pQCD, where $\alpha_{\rm hf}(m_b)\sim 0.18$ and
$\alpha_{\rm hf}(m_c)\sim 0.26$. Just due to such a small value of
$\alpha_{\rm hf}(m_b)$, taken  in bottomonium, small HFS were obtained in
Ref.~\cite{ref.12}, although their wave functions (w.f.) at the origin
gave excellent descriptions of the dielectron widths for $\Upsilon(nS)
(n=1,2,3)$ \cite{ref.13}.

\end{enumerate}

Here we use instead of the Fermi-Breit potential a spin-spin potential
derived using the Field Correlator Method (FCM) \cite{ref.14}, where
relativistic corrections are taken into account and with the mass of a
light quark  $m_n= 5$~MeV $(n=u,d)$ and  $m_s=200$~MeV
for an $s$-quark, the $B, B_s$ mesons can be considered on the
same footing as the $B_c$ mesons.

It can be shown that the HFS are sensitive to the value of
$\Lambda_{\overline{\rm MS}}(n_f)$ taken. Since $\Lambda_{\overline{\rm MS}}$ is
known only for $n_f=5$ and  $\Lambda_{\overline{\rm MS}}$, used for $n_f=3,4$,
varies in a wide range, we make here the assumption, already used in
\cite{ref.4}, that the freezing value of the vector coupling constant
(denoted as $\alpha_{\rm crit}(n_f)$) is the same for $n_f=3,4,5$.

Then it appears possible to obtain a good description of the HFS
for the $B_q$ mesons ($q=n,s,c$) and bottomonium, taking a universal
$\alpha_s(\mu)=0.305(2)$ (with small one-loop corections); its value is
smaller than in Refs.~\cite{ref.5} and \cite{ref.9}.

We also predict the HFS of the as yet undiscovered  $\eta_b(2S)$ and
$\eta_b(3S)$, and the mass of $B_c^*$.

\section{The HF potential in the Field Correlator Method}

The Fermi-Breit potential,
\begin{equation}
 \hat V_{ss}(r) =\ves_1 \cdot \ves_2 \frac{32\pi}{9}\,
 \frac{\alpha_{\rm hf} (\mu)}{\tilde m_1 \tilde m_2}\, \delta^3 (\ver),
\label{eq.01}
\end{equation}
widely used in heavy quarkonia, contains the constituent quark masses
$\tilde m_1$ and $\tilde m_2$, which are very much model-dependent.

In Eq.~(\ref{eq.01}) the strong coupling constant $\alpha_{\rm hf}(\mu)$
may differ from $\alpha_s(\mu)$ (in the $\overline{\rm MS}$ renormalization
scheme) due to higher order perturbative corrections. Here we take the
one-loop corrections from \cite{ref.15},
\begin{equation}
 \alpha_{\rm hf} (\mu) =\alpha_s (\mu) \left[ 1+ \frac{\alpha_s
 (\mu)}{\pi} \rho (n_f)\right],
\label{eq.02}
\end{equation}
where
\begin{equation}
 \rho= \frac{5}{12}\beta_0 - \frac{8}{3} - \frac{3}{4}\ln2.
\label{eq.02a}
\end{equation}

Although higher corrections are small: $\sim 6\%$ for $n_f=3$,
$\sim 3\%$ for $n_f=4$, and $\leq 0.1\%$ for $n_f=5$, still they
are needed to improve the accuracy and have a more self-consistent
picture.

The important role of relativistic corrections, even for the $B_c$
meson, has already been underlined in Refs.~\cite{ref.4} and \cite{ref.9},
and recently in the lattice calculations of the $B_c^*$ mass
\cite{ref.16}, \cite{ref.17}. We take them into account using the
spin-spin potential (without smearing), derived in the FCM
\cite{ref.14}, \cite{ref.18}:
\begin{equation}
 \hat V_{ss}(r) =\ves_1 \cdot \ves_2 \frac{32\pi}{9}\,
 \frac{\alpha_{\rm hf} (\mu)}{\omega_1\omega_2}\, \delta^3 (\ver),
\label{eq.03}
\end{equation}
for which the HFS is
\begin{equation}
 \Delta_{\rm hf} (nS) =
 \frac{8}{9} \frac{\alpha_{\rm hf}(\mu)}{\omega_1\omega_2}|R_n(0)|^2.
\label{eq.04}
\end{equation}
In Eqs.~(\ref{eq.03}) and (\ref{eq.04}) the variables $\omega_1(nS),
\omega_2(nS)$ are the averaged kinetic energies of a quark 1 and a
antiquark 2, which play a role of the dynamical masses:
\begin{equation}
 \omega_1(nS) = \langle \sqrt{\vep^2 + m_1^2} \rangle_{nS}, \quad
 \omega_2(nS) = \langle \sqrt{\vep^2 +m_2^2} \rangle_{nS}
\label{eq.05}.
\end{equation}
The important point is that in Eq.~(\ref{eq.05}) the masses $m_1$
and $m_2$ are well defined; they are the pole masses of $c$ and $b$
quarks (known now with an accuracy of $\sim 70$~MeV for a $b$ quark
and $\sim 100$~MeV for a $c$ quark  \cite{ref.19}). In leading order,
the pole mass does not depend on the number of flavors, while to order
($\alpha_s(\bar m_Q)^2$) it slightly depends on $n_f$ . We take here
$m_1= m_n=5$~MeV for a light quark $(n=u,d)$; $m_s=200$~MeV for an $s$
quark; the pole mass $m_c=1.41$~GeV; $m_b=4.79$~GeV for $n_f=3$ and
$m_b=4.82$~GeV for $n_f=4,5$.

The quantities $\omega_i$ and the w.f. are calculated with the use of the
relativistic string Hamiltonian (RSH), also derived in the FCM \cite{ref.20},
\begin{equation}
 H_0=\frac{\omega_1}{2} +\frac{\omega_2}{2} +\frac{m^2_1}{2\omega_1}+
 \frac{m^2_2}{2\omega_2} +\frac{\vep^2}{2\omega_{\rm red}} +V_{\rm B}(r),
\label{eq.06}
\end{equation}
where the variables $\omega_i$ enter as the kinetic energy
operators. However, if one uses an einbein approximation
\cite{ref.18}, \cite{ref.21} and considers the spin-dependent
potential as a perturbation, then  $\omega_i$ should be replaced by
its matrix elements (m.e.) (\ref{eq.05}).

A simple expression for the spin-averaged mass $M(nS)$ follows from
the RSH \cite{ref.21}:
\begin{equation}
 M(ns)=\frac{\omega_{1}}{2}+\frac{m_1^2}{2\omega_1}+\frac{\omega_{2}}{2}
 +\frac{m_2^2}{2\omega_1} + E_{nS}(\omega_{\rm red}).
\label{eq.07}
\end{equation}
Here, the excitation energy $E_{nS}(\omega_{\rm red})$ depends on the
reduced mass: $\omega_{\rm red}=\frac{\omega_1\omega_2}{\omega_1
+\omega_2}$. The mass formula (\ref{eq.07}) does not contain any
additive constant in the case of bottomonium, while for the $B$ and
$B_s$ mesons a negative (not small) self-energy term, proportional
to $(\omega_q)^{-1}$ ($q=n,s$), has to be added to their masses
\cite{ref.22}.

Then the variables $\omega_i(nS)$, the excitation energy
$E_{nS}(\omega_{\rm red})$, and the w.f. are calculated from the
Hamiltonian (\ref{eq.06}) and two extremum conditions,
$\partial\,M(nS)/\partial\omega_i=0~(i=1,2)$, which are put on the
mass $M(nS)$ \cite{ref.18}:
\begin{eqnarray}\nonumber
&& \left[\frac{\omega_1}{2}+\frac{\omega_2}{2}
+\frac{m_1^2}{2\omega_1} +
 \frac{m^2_2}{2\omega_2}+
 \frac{\bm{p}^2}{2\omega_{\rm red}}+V_{\rm B}(r)
 \right]\varphi_{nS}(r)\\
\label{eq.08}&&= E(nS)\varphi_{nS},
\end{eqnarray}
\begin{equation}
 \omega_i^2(nS)  = m_i^2 -
 \frac{\partial E(nS, \mu_{\rm red})}{\partial \omega_i(nS)},\;  (i=1,2).
\label{eq.09}
\end{equation}

In a Hamiltonian approach the choice of the static potential $V_{\rm
B}(r)$ is of great importance; we take it as a sum of a linear confining
term and the OGE -type term: this property of additivity is well
established now in analytical studies \cite{ref.14} and on the lattice
\cite{ref.23}, \cite{ref.24}:
\begin{equation}
   V_{\rm B}(r)=\sigma\ r +\frac{4\alpha_{\rm B}(r)}{3\ r}.
\label{eq.10}
\end{equation}

For the string tension a conventional value,
$\sigma=0.18$~GeV$^2$, is used here for all mesons.

The main uncertainty comes from the vector coupling $\alpha_V(r)$,
which is taken here from Refs.~\cite{ref.25}, \cite{ref.26} and
denoted as $\alpha_{\rm B}(r)$. Two important conditions have to
be put on the vector coupling:

(i) As in pQCD, it must possess the property of asymptotic freedom
(AF); precisely owing to this property the static interaction
depends on the number of flavors.

(ii) The vector coupling freezes at large distances. The property of
freezing was widely used in phenomenology \cite{ref.4}-\cite{ref.7},
\cite{ref.27} and observed in lattice calculations of the static potential
\cite{ref.23}, \cite{ref.24}.\\

Unfortunately, one cannot use the static potential and the
freezing (critical) constant from lattice studies, where the
latter is found to be significantly smaller than in phenomenology
and background perturbation theory (BPT). There $\alpha_{\rm
B}(\rm crit)=0.58-0.60$ is used (these numbers are close to the
value from \cite{ref.4}). On the lattice, $\alpha_{\rm crit}({\rm
lat})\sim 0.30$ in full QCD ($n_f=3$) \cite{ref.24} and
$\alpha_{\rm crit}({\rm lat})\sim 0.22$ in quenched calculations
\cite{ref.10}, \cite{ref.23} were obtained.

In Eq.~(\ref{eq.10}) the vector coupling $\alpha_{\rm B}(r)$ is
defined via the vector coupling $\alpha_{\rm B}(q^2)$ in momentum
space \cite{ref.26}:
\begin{equation}
 \alpha_{\rm B}(r) =
 \frac{2}{\pi}\int\limits_0^\infty dq\frac{\sin(qr)}{q}\alpha_{\rm B}(q),
\label{eq.11}
\end{equation}
where the vector coupling  $\alpha_{\rm B}(q^2)$ is taken  in two-loop
approximation,
\begin{equation}
 \alpha_{\rm B}(q) =\frac{4\pi}{\beta_0t_{\rm B}}\left(1-\frac{\beta_1}{\beta_0^2}
 \frac{\ln t_{\rm B}}{t_{\rm B}}\right)
\label{eq.12}
\end{equation}
with the logarithm
\begin{equation}
 t_{\rm B}=\frac{q^2+M_{\rm B}^2}{\Lambda_{\rm B}^2},
\label{eq.13}
\end{equation}
containing the vector constant  $\Lambda_{\rm B}(n_f)$, which
differs from the QCD constant $\Lambda_{\overline{\rm MS}}(n_f)$.
The relation between them has been established in
Ref.~\cite{ref.28}:
\begin{equation}
 \Lambda_{\rm B}(n_f)=\Lambda_{\overline{\rm MS}}\exp\left(-\frac{a_1}{2\beta_0}\right),
\label{eq.14}
\end{equation}
with $\beta_0=11 -\frac{2}{3}n_f$ and $a_1=\frac{31}{3}-\frac{10}{9}n_f$.
Therefore the constant $\Lambda_{\rm B}$ is always larger than
$\Lambda_{\overline{\rm MS}}$,
\begin{eqnarray}\nonumber
 \Lambda_{\rm B}^{(5)}&=&1.3656\,\Lambda_{\overline{\rm MS}}^5
 (n_f=5),\;\\ \label{eq.15}
 \Lambda_{\rm B}^{(4)}&=&1.4238\,\Lambda_{\overline{\rm MS}}^4
 (n_f=4),\;\\ \nonumber
 \Lambda_{\rm B}^{(3)}&=&1.4753\,\Lambda_{\overline{\rm MS}}^3 (n_f=3).
\end{eqnarray}

At present, only the QCD constant $\Lambda_{\overline{\rm MS}}(n_f=5)$ is
known with a good accuracy, while for $n_f=3,4$ it is defined  with an
accuracy $\sim 10\%$  \cite{ref.19}). For a given $\Lambda_{\overline{\rm
MS}}(n_f=5)$ one can define $\alpha_{\rm crit}(n_f=5)$. Then, assuming
that the freezing constant is the same for $n_f=3,4$, the QCD  constant,
$\Lambda_{\overline{\rm MS}}$ for $n_f=3,4$, can be calculated.

The mass $M_{\rm B}$ under the logarithm $t_{\rm B}$ is
proportional to $\sqrt\sigma$ and for $\sigma=0.18$~GeV${}^2$  its
value is equal to $M_{\rm B}=1.0\pm 0.05$~GeV \cite{ref.29}.

The critical constants $\alpha_{\rm B}(\rm crit)=\alpha_{\rm B}(q^2=0)$
in momentum space and $\alpha_{\rm B}(r\rightarrow \infty)$ in coordinate
space are proved to be equal \cite{ref.26}.

Our calculations give small relativistic corrections for bottomonium:
$\omega_b(1S) - m_b \sim 190$~MeV ($\sim 4\%$) and $\sim 7\%$ for the $2S$
and $3S$ states. It is of interest to notice that the relativistic correction
to the $b-$quark mass is even smaller in the $B_c$ meson,  $\omega_b(1S)
-m_b\sim 83$~MeV ($\sim 2\%$), while for a $c-$quark the difference
$\omega_c(1S)- m_c \sim 250$~MeV is already $\sim 20\%$. The values of
$\omega_q(1S)-m_q$ are given in Table~\ref{tab.1} for the $B_q$ mesons
($q=n,s,c$).

\begin{table}
\caption{\label{tab.1}The kinetic energies $\omega_q(1S)~(q=n,s,c)$ and
$\omega_b(1S)$ (in MeV) for the static potential $V_{\rm B}(r)$
(\ref{eq.10}) with $n_f=4$ and $\alpha_{\rm crit}=0.58$}
\begin{center}
\begin{tabular}{cccc}
\hline\hline
    \quad    Meson \quad    &       $  B$  &     $B_s$   &    $B_c$\\
\hline
    $ m_q $       &        5    &    200  &       1410 \\
 $\omega_q  -m_q$ &       611  &    489   &     248\\
 $\omega_b$       &        4805 &    4825 &      4888\\
 $\omega_b -m_b $ &    25      &  25     &  83 \\
\hline\hline
\end{tabular}
\end{center}
\end{table}

It is convenient to introduce the ratio  $g_{B_q}$,
\begin{equation}
 g_{B_q}(nS) = \frac{|R_n(0)|^2}{\omega_1(nS)\omega_2(nS)},
\label{eq.16}
\end{equation}
which directly enters the HFS (\ref{eq.04}) and appears to be weakly
dependent on small variations of the masses $m_1$ and $m_2$, which are
compatible with a good description of the meson spectrum.

The w.f. at the origin are sensitive to the values of
$\Lambda_{\rm B}(n_f)$ (in \cite{ref.30} only the case with
$n_f=3$ was considered). However, if the same freezing value of
the coupling constant is taken for $n_f=3,4,5$, then the
differences between the w.f. at the origin for $n_f=3$ and $n_f=4$
are $\leq 3\%$.

We use here two values for $\alpha_{\rm crit}$: $\alpha_{\rm crit}=0.580$
and 0.604, for which  corresponding values of $\Lambda_{\rm B}(n_f)$,
$\Lambda_{\overline{\rm MS}}(n_f)$  are given below in Eq.~(\ref{eq.17}).

\begin{table}
\caption{\label{tab.2} The ratios $g_{B_q}$ (\ref{eq.09}) (in GeV)
and $|R_1(0)|^2$ (in GeV${}^3$) for the $B_q(1S)$ mesons
($\alpha_{\rm crit}=0.58, n_f=4$)}
\begin{center}
\begin{tabular}{lccc}
\hline\hline
           &     $ B  $ &        $ B_s $  &      $ B_c$\\
\hline
$|R(0)|^2$ &      ~~0.477~~   &     ~~0.551~~   &   ~~1.669~~\\
$g_{B_q}$  &       ~~0.161~~ &     ~~0.166~~  &     ~~0.205~~\\
\hline\hline
\end{tabular}
\end{center}
\end{table}

In bottomonium the difference between $g_b$ for $n_f=4$ and
$n_f=5$ appears to be larger, $\sim 10\%$ (see Table~\ref{tab.3}), where in
both cases $\alpha_{\rm crit}=0.604$ is used.

\begin{table}
\caption{\label{tab.3} The ratios $g_b(nS)$ (in GeV) and $|R_n(0)|^2$ (in
GeV${}^3$) for the $1S$, $2S$, and $3S$ bottomonium states with
$\alpha_{\rm crit}=0.604$ and for $n_f=4,5$}
\begin{center}
\begin{tabular}{lccc}
\hline\hline
                           & 1S &        2S &        3S\\
\hline
 $|R_n(0)|^2 (n_f=5)$  &     6.476 &        ~~~3.398 () &   ~~2.682()\\
  $g_b  (n_f=5) $ &         0.258 &        ~~~~0.134 (1)  &   ~~~~0.105(1)\\
 $|R_n(0)|^2  (n_f=4)$&    5.668  &        3.126 &    2.508\\
  $  g_b   (n_f=4) $&      0.230&          0.127  &    0.100\\
  $ \Delta_{hf}(n_f=5)  $&      70.3 &          36.3  &    28.4 \\
\hline\hline
\end{tabular}
\end{center}
\end{table}

For the values of $\alpha_{\rm crit}(n_f)=0.604 (0.58)$ and
$\Lambda_{\overline{\rm MS}}(n_f=5)=0.245 (0.236)$~GeV, the following
vector constants: $\Lambda_{\rm B}(n_f=3)=0.40 (0.389)$~GeV; $\Lambda_{\rm
B}(n_f=4)=0.372 (0.360)$~GeV, $\Lambda_{\rm B}(n_f=5)=0.335 (0.323)$~GeV
are obtained. Then from the relation (\ref{eq.14}) we have
\begin{eqnarray}
\nonumber
 \Lambda_{\overline{\rm MS}}(n_f=3)&=&271 (264)~{\rm MeV},\;\\ 
 \Lambda_{\overline{\rm MS}}(n_f=4)&=&261 (253)~{\rm MeV},\;\\ \nonumber
 \lambda_{\overline{\rm MS}}(n_f=5)&=&245 (236)~{\rm MeV}.
\label{eq.17}
\end{eqnarray}
For $n_f=5$ it gives $\alpha_s(M_Z)({\rm two-loop})=0.1194$ for
$\alpha_{\rm crit}=0.604$ and $\alpha_s(M_Z)=0.1188$ for $\alpha_{\rm
crit}=0.58$; both numbers agree with the world averaged value,
$\alpha_s(M_Z)=0.1176\pm 0.0020$ \cite{ref.19} within its error bar.

\section{Results}

The experimental error in the HFS
\begin{equation}
 \Delta_{\rm hf}(b\bar b) =M(\Upsilon(9460))- M(\eta_b)=69.9\pm 3.1~{\rm MeV}
\label{eq.18}
\end{equation}
is small, $\pm 3$~MeV, and it is even smaller, $\leq 1$~MeV, for the mass
differences $M(B^*)-M(B)$, $M(B_s^*)-M(B_s)$ \cite{ref.19}. Therefore,
we expect that the coupling constant $\alpha_{\rm hf}$ can be extracted
with a good accuracy from these data.

Notice that from Eq.~(\ref{eq.02}) and the value $\alpha_s(\mu)=0.305(2)$
one obtains the following coupling constants with one-loop corrections:
$\alpha_{\rm hf}(n_f=5)=\alpha_s(\mu)=0.305(2)$, $\alpha_{\rm hf}(n_f=4)=
0.314(2)$, and $\alpha_{\rm hf}(n_f=3)= 0.323(2)$. Precisely these values
are used in our analysis.

For  $g_b=0.230$~GeV ($n_f=4$) and $g_b=0.258$~GeV
($n_f=5$) (see Table~\ref{tab.3}) and taking $\alpha_{\rm
hf}(n_f=5)=\alpha_s(\mu)=0.305(2)$ and $\alpha_{\rm hf}(n_f=4)=0.314(2)$,
we obtain $\Delta_{\rm hf}(b\bar b)= 64.2(4)$~MeV ($n_f=4$) and
70.0(4)~MeV for $n_f=5$. The difference between them, $\sim 10\%$, is
not small and one may conclude that the HFS in bottomonium is in full
agreement with the BaBar data \cite{ref.1}, \cite{ref.2} only for $n_f=5$.

For the $2S$ and $3S$ bottomonium states the difference between the
cases with $n_f=4$ and $n_f=5$ is small; their HFS coincide within
2~MeV. Therefore we predict that $\Delta(b\bar b)$ is equal to 36(2)~MeV
and 27(2)~MeV, respectively, for the $3S$ and $2S$ states.

For the $B_q$ mesons, the cases with $n_f=3,4$ have been
considered and in both cases agreement with experiment is obtained
(see Table~\ref{tab.4}); it happens because the by $\sim 3\%$
smaller value of $g_{B_q}$ is compensated by the $\sim 3\%$ larger
value of $\alpha_{\rm hf}(n_f=3)$. Thus for $B$ and $B_s$ the
number of flavors cannot be fixed, if only data on the HFS (and
the spectrum) are fitted; therefore some additional information,
like decay constants, is neededd to fix  $n_f$.

\begin{table}[h]
\caption{\label{tab.4} The HFS (in MeV) of the $B_q$ mesons  with
$\alpha_s(\mu)=0.307$ ($\alpha_{\rm hf}(n_f=4)=0.316$, $\alpha_{\rm
hf}(n_f=3)=0.324 )$}
\begin{center}
\begin{tabular}{lllll}
\hline\hline
                    & $B$   &     $B_s$ & $B_c(1S)$ & $B_c(2S)$ \\
\hline
 $\Delta_{\rm hf}(n_f=4)$ &  45.2           &  46.5      &   57.8 & 38.0 \\
 $\Delta_{\rm hf}(n_f=3)$ &  45.4     &        $ 46.1 $ &   ~57.2~ & 37.4  \\
 $\Delta_{\rm hf}$(exp) & $45 .78\pm 0.35$ & $46.5\pm 1.25$ & absent & absent\\
\hline\hline
\end{tabular}
\end{center}
\end{table}

From our calculations we predict the following masses of the triplet
and singlet $c\bar b(2S)$ states: $M(B_c^*(2\,{}^3S_1))=6901(5)$~MeV  
and $M(B_c(2\,{}^1S_0))= 6865(5)$~MeV, which are calculated in
single-channel approximation.

Notice that the extracted strong coupling constant,
$\alpha_s(\mu)=0.305(2)$, is smaller than the one used in
Refs.~\cite{ref.5} and \cite{ref.9}. The renormalization scale,
$\mu\sim 1.6-1.65$~GeV, corresponding to this coupling, is rather
large but agrees with the existing interpretation of the spin-spin
potential as dominantly perturbative, thus partly justifying the
use of the $\delta^3(\ver)$-function.

\section{Conclusion}

Our study of the HFS is performed assuming that the freezing value
of the coupling constant is the same for $n_f=3,4,5$ and considering
$\alpha_{\rm crit}=0.58$ and 0.60.  The calculated HFS of the $B$ and
$B_s$ mesons are in good agreement with experiment for both freezing
constants. It happens that the HFS for $n_f=3$ and $n_f=4$ coincide with
each other (within 0.5~MeV), if the one-loop correction is taken into
account in $\alpha_{\rm hf}(n_f)$. The HFS, averaged over two results,
are $M(B^*) - M(B)= 45.3(2)$~MeV, $M(B_s^*)-M(B_s)=46.5(3)$~MeV, and
$M(B_c^*)-M(B_c)=57.5(10)$~MeV. The latter number gives the mass of the as
yet unobserved $B_c^*$ meson, $M(B_c^*)=6.334\pm 1$~MeV. For excited
$B_c(2S)$ states we predict the masses: $M(B_c^*(2S)=6901(5)$~MeV and
$M(B_c(2S)=6865(5)$~MeV, which are calculated neglecting open channels.

In bottomonium the choice $n_f=5$, where $\Delta_{\rm hf}(b\bar b)=70.0(4)$~MeV,
is considered preferable, since for $n_f=4$ the HFS is smaller, equal to
64.2(4)~MeV .

For the $2S$ and $3S$ bottomonium states, the calculated HFS are 36(1)~MeV
and 27(1)~MeV, respectively.

The extracted coupling, $\alpha_s(\mu)=0.305(2)$, is smaller than
in many other analyses with a universal coupling; it determines
the characteristic scale of the spin-spin interaction, being $\mu
\sim 1.6-1.65$~GeV. Knowledge of this scale can help to better
understand the behavior of the spin-spin potential at small
distances.

\begin{acknowledgments} This work is supported by the Grant
RFFI-09-02-00629 a.
\end{acknowledgments}

\end{document}